\documentclass{ws-procs9x6-cpt19}
\usepackage{bm}
\begin{document}

\newcommand{\refeq}[1]{(\ref{#1})}
\def\etal {{\it et al.}}

\title{Pulsar tests of the gravitational Lorentz violation}

\author{Lijing Shao}

\address{Kavli Institute for Astronomy and Astrophysics, Peking University, \\
Beijing 100871, China}

\begin{abstract}
    Pulsars are precision celestial clocks. 
    When being put in a binary, the ticking conveys the secret of underlying spacetime geometrodynamics. 
    We use pulsars to test if the gravitational interaction possesses a tiny deviation from Einstein's General Relativity (GR).
    In the framework of Standard-Model Extension (SME),
    we systematically search for Lorentz-violating operators cataloged by
    (a) the minimal couplings of mass dimension 4, 
    (b) the CPT symmetry of mass dimension 5, and 
    (c) the gravitational weak equivalence principle (GWEP) of mass dimension 8.
    No deviation from GR was found yet.
\end{abstract}

\bodymatter

\section{Introduction}

Pulsars are magnetized rotating neutron stars (NSs). 
Their stable rotation forms precision celestial clocks across the sky. 
Though being thousands of light years away, we can use the technology of pulsar timing to obtain accurate physical inferences.\cite{pulsar:gravity}
For a binary system, the orbital motion is determined by the gravitational interaction. Thus, a binary pulsar provides a fundamental way to look into whether Einstein's General Relativity (GR) correctly describes the gravity.

In GR, spacetime is a 4-dimensional differential manifold and 
the tangent space of every single point has the symmetry of local Lorentz invariance (LLI). 
In the searches for a fundamental quantum-gravity theory, LLI was questioned. 
The so-called Standard-Model Extension (SME) presents a practically convenient way to systematically investigate the possibility of Lorentz violation.\cite{sme:theory} 
Binary pulsars were proposed to be excellent laboratories to study the gravity sector of the SME.\cite{sme:gravity} 
In this contribution, we present a short summary on what have been achieved along this line of investigation.\cite{sme:pulsar} 

\section{The gravity sector of the Standard-Model Extension}

In the gravity sector of SME, generic Lorentz-violating operators are added to the Lagrangian of GR. 
Being sorted by these operators' mass dimension, the Lagrangian in the linearized limit is,\cite{sme:theory, sme:gravity}
\begin{equation}
    {\cal L} = {\cal L}_{\rm GR} + {\cal L}_{\rm SME}^{(4)} + {\cal L}_{\rm SME}^{(5)} + \cdots \,,
\end{equation}
where,
\begin{eqnarray}
    {\cal L}_{\rm GR} &=& - \frac{1}{32\pi G} h^{\mu\nu} G_{\mu\nu} + \frac{1}{2} h_{\mu\nu} T^{\mu\nu}_{\rm matter} \,, \label{eq:GR} \\
    {\cal L}^{(4)}_{\rm SME} &=& \frac{1}{32\pi G} \bar s^{\mu\kappa} h^{\nu\lambda} {\cal G}_{\mu\nu\kappa\lambda} \,, \label{eq:sme4} \\
    {\cal L}^{(5)}_{\rm SME} &=& - \frac{1}{128\pi G} h_{\mu\nu} q^{\mu\rho\alpha\nu\beta\sigma\gamma} \partial_\beta R_{\rho\alpha\sigma\gamma} \,. \label{eq:sme5}
\end{eqnarray}
In the above expressions, $\bar s^{\mu\kappa}$ and $q^{\mu\rho\alpha\nu\beta\sigma\gamma}$ are Lorentz-violating vacuum expectation values that are resulted from the spontaneous symmetry breaking of corresponding dynamical tensor fields.\cite{sme:theory, sme:gravity}

\section{Pulsar tests}

Lorentz-violating operators were constrained by the precision pulsar timing; see the annually updating {\it Data Tables for Lorentz and CPT Violation}\cite{sme:table} for a comprehensive summary.

\subsection{The minimal couplings}

The Lagrangian \refeq{eq:sme4} gives the mass dimension 4 couplings in the SME. They are the lowest-order operators.\cite{sme:gravity} Because of the tensorial nature of $\bar s^{\mu\kappa}$, multiple binary pulsars with different sky location and different orbital orientations are extremely powerful to break parameter degeneracy when constraining $\bar s^{\mu\kappa}$'s various components. The first study of pulsars in the gravity sector of SME constructed 27 independent tests from 13 pulsar systems. The $\bar s^{{\rm T}k}$ and $\bar s^{jk}$ $(j,k={\rm X,Y,Z})$ components are jointly constrained to the levels of ${\cal O} \left(10^{-9}\right)$ and ${\cal O} \left(10^{-11}\right)$ respectively.\cite{sme:pulsar} In addition, by using the boost brought by the systematic velocity between binary pulsars and the Solar System, $\bar s^{\rm TT}$ is constrained to be smaller than ${\cal O} \left(10^{-5}\right)$.\cite{sme:pulsar}

\subsection{The CPT symmetry}

The Lagrangian \refeq{eq:sme5} breaks the CPT symmetry.\cite{sme:gravity} The abnormal acceleration introduced is proportional to $\sim qv \times Gm_1 m_2  / r^3$ where ${\bf v}$ is the relative velocity between the attractor and the receiver of the gravity. Due to the appearance of $v$, static experiments in ground-based laboratories are extremely disadvantageous to put constraints. In the contrast, binary pulsars have a significant relative velocity $v\sim 10^3 \, {\rm km\,s}^{-1}$. In the 
``maximal-reach'' approach where different components are one by one considered nonvanishing, $q^{\mu\rho\alpha\nu\beta\sigma\gamma}$ are constrained to be less than ${\cal O}\left(10^0\right)$--${\cal O}\left(10^1\right)$\,m.\cite{sme:pulsar}

\subsection{The gravitational weak equivalence principle}

In the above subsections, only the quadratic terms of $h_{\mu\nu}$ are
considered in the Lagrangian. If the cubic couplings are taken into account,
new phenomena appear.\cite{sme:gravity, sme:pulsar} The leading terms in this scenario read,
\begin{equation}
  \label{eq:sme8}
    \frac{\sqrt{-g}}{16\pi G} k^{(8)}_{\alpha\beta\gamma\delta\kappa\lambda\mu\nu\epsilon\zeta\eta\theta} R^{\alpha\beta\gamma\delta} R^{\kappa\lambda\mu\nu} R^{\epsilon\zeta\eta\theta} \subset {\cal L}^{(8)}_{\rm SME}  \,.
\end{equation}
Such a Lagrangian introduces compactness-dependent accelerations that violate
the gravitational weak equivalence principle (GWEP) for self-gravitating
objects. The more compact the object, the larger the abnormal acceleration. NSs
with compactness $\sim GM/Rc^2 \simeq 0.2$ (compared with $<10^{-26}$ for
terrestrial experiments) are excellent objects to probe the GWEP-violating signals. 
The first empirical study of Eq.\ \refeq{eq:sme8} was conducted with pulsar timing, that sets limit on $k^{(8)}_{\alpha\beta\gamma\delta\kappa\lambda\mu\nu\epsilon\zeta\eta\theta}$ at the level of ${\cal O} \left( 10^2 \right) \, {\rm km}^4$ in the maximal-reach approach.\cite{sme:pulsar}

\section{Discussion}

In this contribution, pulsar tests of the gravitational Lorentz violation are
reviewed concisely in the framework of SME. These tests are related to the minimal couplings, the CPT symmetry and the GWEP. No violation was discovered. Complementary tests in searching for tiny deviations from GR were performed in the alternative metric-based parametrized post-Newtonian (PPN) formalism.\cite{pulsar:ppn}

In principle, NSs are strong-field objects\cite{pulsar:gravity} and a strong-field version of SME or PPN should be applied. However, generic frameworks with strong-field objects are still under development. Some strong-field phenomena were studied in specific classes of theories, e.g., the scalar-tensor gravity.\cite{strong:field} In this sense, tests presented here are {\it effective} strong-field counterparts. Generally, these tests are enhanced due to the strong fields. Therefore, we consider the limits reviewed here are conservative ones.

With more pulsars to be discovered and more precision measurements to be made by the Five-hundred-meter Aperture Spherical Telescope (FAST) and the Square Kilometre Array (SKA)\cite{ska:fast} in the near future, better limits are guaranteed. If Nature possesses deviations from the Lorentz symmetry, a discovery will change our understanding of the physical world forever.

\section*{Acknowledgments}
This work was supported by the Young Elite Scientists Sponsorship Program by
the China Association for Science and Technology (2018QNRC001). It was
partially supported by the National Natural Science Foundation of China
(11721303), 
and the Strategic Priority Research Program of the Chinese Academy of
Sciences through the grant No. XDB23010200.

\end{document}